\documentclass[12pt]{article}
\usepackage{amssymb}
\usepackage{graphicx}
\usepackage{latexsym,cite}
\newcommand{\eq}{\begin{equation}}
\newcommand{\en}{\end{equation}}
\newcommand{\be}{\begin{equation}}
\newcommand{\ee}{\end{equation}}
\newcommand{\qe}{\end{equation}}
\newcommand{\ear}{\begin{eqnarray}}
\newcommand{\eqa}{\begin{eqnarray}}
\newcommand{\ba}{\begin{eqnarray}}
\newcommand{\ea}{\begin{eqnarray}}
\newcommand{\rae}{\end{eqnarray}}
\newcommand{\ena}{\end{eqnarray}}
\newcommand{\beq}{\begin{equation}}
\newcommand{\eeq}{\end{equation}}
\newcommand{\bea}{\begin{eqnarray}}
\newcommand{\eea}{\end{eqnarray}}
\newcommand{\Z}{\mathbb{Z}}
\newcommand{\N}{\mathbb{N}}
\newcommand{\smt}{\sum_{n=0}^\infty}
\def\de{\partial}
\begin{document}
\begin{titlepage}
\vskip0.5cm
\begin{flushright}
DFTT 04/05\\
IFUP-TH 2005-04\\
DIAS-STP-05-01 \\
\end{flushright}
\vskip0.5cm
\begin{center}
{\Large\bf Comparing the Nambu-Goto string with LGT results}
\end{center}
\vskip1.3cm
\centerline{
M. Caselle$^{a}$, M. Hasenbusch$^{b}$
 and M. Panero$^{c}$}
 \vskip1.0cm
 \centerline{\sl  $^a$ Dipartimento di Fisica
 Teorica dell'Universit\`a di Torino and I.N.F.N.,}
 \centerline{\sl via P.Giuria 1, I-10125 Torino, Italy}
 \centerline{\sl
e--mail: \hskip 1cm
 caselle@to.infn.it}
 \vskip0.4 cm
 \centerline{\sl  $^b$  Dipartimento di Fisica dell'Universit\`a di Pisa
 and I.N.F.N.,}
   \centerline{\sl Largo Bruno Pontecorvo 3, 
                  I-56127 Pisa,
                        Italy}
 \centerline{\sl
e--mail: \hskip 1cm
 Martin.Hasenbusch@df.unipi.it}
\vskip0.4 cm
 \centerline{\sl  $^c$ School of Theoretical Physics,
Dublin Institute for Advanced Studies,}
 \centerline{\sl
                  10 Burlington Road, Dublin 4,
                              Ireland}
 \centerline{\sl
e--mail: \hskip 1cm
 panero@stp.dias.ie}
 \vskip1.0cm

\begin{abstract}

We 
discuss a way
to evaluate 
the full prediction for the interquark potential which is expected from the effective Nambu-Goto string model. We check the correctness of
the prescription 
reproducing  the results obtained with the zeta function regularization
for the first two perturbative orders. We compare 
the predictions with existing 
Monte Carlo
data for the (2+1) dimensional $\Z_2$, $SU(2)$ and $SU(3)$ gauge theories: 
in the low temperature regime, we find good agreement for large enough interquark distances, but an increasing mismatch between theoretical predictions and numerical results is observed as shorter and shorter distances are investigated. 
On the 
contrary, at high temperatures (approaching the deconfinement transition 
from below) 
a remarkable agreement between 
Monte Carlo 
data and 
the expectations from the Nambu-Goto effective string is observed for a wide range of interquark distances.

\end{abstract}
\end{titlepage}

\setcounter{footnote}{0}
\def\thefootnote{\arabic{footnote}}

\section{Introduction}
\label{sect1}
Working out
an effective
description for the infrared behaviour of the interquark potential is one of the
major challenges of modern lattice gauge theory \cite{lsw,alvarez81,df83}. 
In these last years, much effort has been devoted in trying to identify the nature of 
an effective string model 
that reproduces the results of high precision numerical studies 
of the confining potential in pure gauge models \cite{lw01}--\cite{jp04}.
Despite these efforts, a clear answer 
has not been obtained 
yet. The simplest, and most natural, candidate 
for an effective string picture is the Nambu-Goto model 
\cite{Nambu:1970,Goto:1971ce,Nambu:1974zg}:
numerical tests of the predictions of this model at the first leading orders, however,
did not lead to 
conclusive results. As a matter of fact, the data obtained for different 
lattice gauge theories at short interquark distances $R$ disagree with the Nambu-Goto
predictions
\cite{chp04,cpr04,Juge:2002br,Juge:2003ge,jkm03}, whereas at large distances 
the agreement is remarkably good~\cite{chp03,chp04,cpr04}.

In the setting where one considers a confined, static, quark-antiquark ($Q\bar{Q}$) pair at a finite temperature $T$, a possible way to derive effective predictions from the Nambu-Goto model consists in working out a perturbative expansion in powers of $(\sigma RL)^{-1}$, a dimensionless quantity which depends on the 
interquark distance $R$, on the system size in the compactified direction $L$ (which is inversely proportional to the temperature) and on the string tension $\sigma$. 
As it is expected, the leading, non-trivial (LO) contribution emerging in this expansion describes the asymptotic free string picture, whose degrees of freedom are independent, massless bosons associated with string vibrations along the transverse directions, 
whose partition function can be easily evaluated by means of standard conformal field theory techniques. The contribution at the next-to-leading order (NLO) can also be evaluated analytically \cite{df83}, but treating further terms in such a perturbative expansion would be much more difficult. 
On the other hand, comparison of numerical results with the next-to-leading-order predictions can be tricky, as it is 
difficult to 
judge
if the mismatch observed, for instance, at short distances and low temperatures could be interpreted as an effect due to the 
NLO truncation, or if --- \emph{vice versa} --- the agreement which is found at high temperatures could be spoiled by higher orders (see discussion in section 5 of~\cite{chp03}). 

Here we address this problem by estimating 
the full partition function of the 
Nambu-Goto effective string model, 
relying on the fact that formal quantization of this model is possible, 
and the spectrum can be obtained analytically \cite{Arvis:1983fp}. 
This procedure allows for a comparison with the existing 
Monte Carlo data of the interquark potential that is not biased by
the truncation effects discussed above.

The results of this comparison can be summarized as  follows:
\begin{itemize}
\item
The mismatch between numerical data and NLO predictions which is observed at
short distances is confirmed (and even enhanced) if one considers the full string prediction.\footnote{The authors of ref.~\cite{jkm03} reported a similar result for the effective conformal charge and the first excited string state.}
\item
On the 
contrary, 
at higher temperatures,
approaching the deconfinement transition from below, 
a remarkable agreement (which is particularly striking for the case of $SU(2)$ gauge group) between 
Monte Carlo data and 
full Nambu-Goto expectations 
is found.
\end{itemize}

These observations suggest that the 
Nambu-Goto action 
does not
provide a correct description of the 
interquark potential at 
short distances, where
the problem is still open and
an alternative effective
model is probably needed,
in agreement with the results which were first presented in \cite{Juge:2002br,jkm03}.
At the same time, however, these observations also suggest that this 
effective string action can be considered as a
reliable (and better than the pure ``free string picture'') description
of the interquark potential 
at large distances.
As the finite-temperature deconfinement transition is approached we observe a crossover between 
this ``universal'' Nambu-Goto behaviour and a new behaviour which is different in the $\Z_2$ and $SU(2)$ cases and keeps into account the fact the Nambu-Goto action would lead to the wrong critical indices at the deconfinement point.

This paper is organized as follows. In section 
\ref{nambugotostringsect}, after recalling the basics about the 
Nambu-Goto model, 
we work out the predictions of interest 
for the effective string scenario, using the 
perturbative approach and evaluating the full partition 
function in terms of the string spectrum. We also  
comment about the convergence bounds that 
naturally arise in the latter approach. 
In section \ref{comparisonsect}, 
the predictions of the full string partition function 
are compared with existing 
Monte Carlo data for 
$\Z_2$, $SU(2)$ and $SU(3)$ LGT in $(2+1)$ dimensions. Finally, in section 
\ref{conclusionsect}
we summarize the results and provide our conclusions.
Calculations 
displaying the matching between the full partition 
function and the known perturbative results at leading and 
next-to-leading order are reported in the Appendix.

\section{The Nambu-Goto string} 
\label{nambugotostringsect}

In this section, we report some results concerning the  
Nambu-Goto string model   
\cite{Nambu:1970,Goto:1971ce,Nambu:1974zg} which will be useful to our analysis. We refer the reader to \cite{chp03} for a more detailed derivation.
Before presenting the theoretical predictions of this model, we would like to remark the fact that the IR description in terms of an effective string between the sources is eventually expected to break down at some finite distance scale: as a matter of fact, at short distances it is reasonable to expect that the behaviour of the interquark potential gets a contribution from the gluonic modes inside the quark-antiquark flux tube \cite{Juge:2003ge} --- whose role is irrelevant in the picture of a long enough, string-like, flux tube.

The   
action of the Nambu-Goto model is proportional to the area of the string 
world-sheet:
\eq
S=\sigma\int_0^{L}d\tau\int_0^{R} d\varsigma\sqrt{g}\ \ ,\label{action}
\en
where $g$ is the determinant of the two--dimensional metric induced on
the world--sheet by the embedding in $R^d$  
and $\sigma$ is the string tension,
which appears as a parameter of the effective model.
\par
Eq.~(\ref{action}) is invariant with respect to reparametrization and Weyl
transformations, and a possible choice for quantization of the effective model
is the ``physical gauge'' (see \cite{alvarez81} for more details) in which $g$ becomes a function of the transverse displacements of the string world-sheet only. The 
latter (which can be denoted as $X^i(\tau,\varsigma)$) are required to satisfy the boundary conditions relevant to the problem --- in the present case, the effective string world-sheet associated with a two-point Polyakov loop correlation function obeys periodic b.c. in the compactified direction and Dirichlet b.c. along the interquark axis direction:
\be
X^i(\tau+L,\varsigma)=X^i(\tau,\varsigma); \hskip 1cm X^i(\tau,0)=X^i(\tau,R)=0\ \ .
\en
This gauge fixing implicitly assumes that the world-sheet surface is a single-valued function of $(\tau,\varsigma)$, i.e. overhangs, cuts, or disconnected parts are excluded. 
It is well known that rotational symmetry of this model is broken at a quantum level because of the Weyl anomaly, unless the model is defined in the critical dimension $d=26$. However, this anomaly is known to be vanishing at large distances \cite{olesen}, and this suggests to use the ``physical gauge'' for
an IR, effective string description also for $d \neq 26$.

Here and in the following, we restrict our attention to the $d=2+1$ case, which is particularly simple, as there is only one transverse degree of freedom ($X$).
In the physical gauge, eq. (\ref{action}) takes the form: 
\eq
S[X]=\sigma\int_0^L d \tau \int_0^R d \varsigma
\sqrt{1+(\de_\tau X)^2+(\de_\varsigma X)^2} \;\;.
\label{squareroot}
\en

This action describes a non-renormalizable, interacting 
QFT in two dimensions; the associated partition function is expected to encode an effective description for this sector of the gauge theory, providing a prediction for the VEV of the two-point Polyakov loop correlation function: 
\eq
\label{partitionfunction}
\langle P^\dagger (R) P(0) \rangle = Z = \int \mathcal{D} X e^{-S[X]} \;\;.
\en
Quantum corrections
beyond the classical solution (which corresponds to a flat string world-sheet surface) can be studied in two different ways.

\subsection{Perturbative expansion of the partition function}\label{perturbativeexpansionsubsect}

Despite the fact that the Nambu-Goto string theory is non-renormalizable, one can address the study of the effective model in a perturbative way, expanding the square root appearing in eq.~(\ref{squareroot}) in powers of the dimensionless quantity $(\sigma RL)^{-1}$: this approach is expected to be consistent with the fact that the effective theory holds in the IR regime of the confined phase. 

Going to dimensionless world-sheet coordinates:
$\xi_1=\tau/L$, $\xi_2=\varsigma/R$ and with a global rescaling for the $X$ field: $\phi = \sqrt{\sigma} X$, 
eq.~(\ref{squareroot}) can then be expanded in a natural way in a series of terms associated with different powers of $(\sigma RL)^{-1}$; in particular, the first few terms read:
\eq
\label{SexpansionNLO}
S= \sigma L R+\frac{1}{2}  \int_0^1 d \xi_1 \int_0^1 d \xi_2
\left(\nabla \phi \right)^2 - \frac{1}{8\sigma L R} \int_0^1 d \xi_1 \int_0^1 d \xi_2
\left[ \left(\nabla \phi\right)^2 \right]^2 + O\left( (\sigma RL )^{-2} \right)\;\;,  
\en
where:
\eq
\left(\nabla \phi\right)^2=
\frac{1}{2u}\left(\frac{\de \phi}{\de \xi_{1}}\right)^{2}+2u
\left(\frac{\de \phi}{\de \xi_{2}}\right)^{2}
\en
and:
\eq
u=\frac{L}{2R} \;\;. 
\en
Apart from the trivial, classical contribution given by the first term appearing on the r.h.s. of eq.~(\ref{SexpansionNLO}) --- the second term of the expansion shows that the leading order quantum correction is nothing but the CFT of a free, massless bosonic field, while the 
subsequent
$O((\sigma RL )^{-1})$ contribution (which is quartic in $\phi$) encodes string self-interaction.

Discarding the string self-interaction term, the LO approximation of the partition function can be evaluated analytically:
\eq
Z^{LO}(L,R)=e^{-\sigma RL} \cdot Z_{1}\  \;\;, 
\label{zlo}
\en
where $Z_{1}$ encodes the (regularized) contribution of the Gaussian fluctuations:
\eq
Z_{1}=\frac{1}{\eta(iu)} 
\label{z1}
\en
and it is expressed in terms of Dedekind's $\eta$ function:
\eq
\eta(\tau)=q^{1\over24}\prod_{n=1}^\infty(1-q^n)\hskip0.5cm
;\hskip0.5cmq=e^{2\pi i\tau}~~~,
\label{eta}
\en

\vskip 0.3cm 
On the other hand, keeping into account the string self-interaction term, the
resulting approximation for $Z$ at the next-to-leading order reads~\cite{df83}:
\eq
Z^{NLO}(L,R)=e^{-\sigma RL} \cdot Z_{1} \cdot \left\{ 1+ \frac{ \pi^2 L}{1152 \sigma R^3} \left[ 2E_4 \left( i u \right) - E_2^2 \left( i u \right) \right] \right\} \;\;, 
\label{znlo}
\en
where $E_2$ and $E_4$ are the Eisenstein functions:
\eqa
E_2(\tau)&=&1-24\sum_{n=1}^\infty \sigma_1(n) q^n \label{eisenstein2}\\
E_4(\tau)&=&1+240\sum_{n=1}^\infty \sigma_3(n) q^n \label{eisenstein4}\\
q&=& e^{2\pi i\tau} \;\;, 
\ena
where $\sigma_1(n)$ and $\sigma_3(n)$ are, respectively, the sum of all
divisors of $n$ (including 1 and $n$), and the sum of their cubes.

\subsection{The string spectrum} 
\label{stringspectrumsubsect}

The spectrum of the Nambu-Goto string with length $R$ and fixed ends can be obtained through formal canonical quantization \cite{Arvis:1983fp}, and in $d=2+1$ dimensions it reads:
\eq
\label{arvisspectrum}
E_n(R) =\sigma R\sqrt{1+\frac{2\pi}{\sigma R^2}
\left( n- \frac{1}{24}\right) }\hskip0.5cm
,\hskip0.5cm n \in \N ~~~.
\en
The full partition function of 
the Nambu-Goto string can thus be written as
--- see also~\cite{lw04}: 
\eq
Z=\sum_{n=0}^{\infty} w_n e^{-E_n L}
\label{wholeng}
\en
with integer weights $w_n$
accounting for level multiplicities. In particular, in $d=2+1$ the latter 
are nothing but $P(n)$: the number of partitions of $n$. 
In fact, since the $w_n$ coefficients do not depend on $R$, 
they can be evaluated in the large $R$ limit, where the theory 
eventually reduces to a $c=1$ CFT in two dimensions. 
The Hilbert space of such a theory is well known
and the multiplicity of states with energy $E_n$ is given by the number 
of ways 
one can 
combine the Virasoro generators 
of the 2d conformal algebra 
to obtain a field of conformal weight $n$ --- which is precisely $P(n)$.

In principle, eq.~(\ref{wholeng}) can provide a tool to test the validity range of the effective Nambu-Goto string picture, disentangling truncation errors which affect the perturbative expansion of $Z$, from physical effects related to the eventual breakdown of this effective string description at some scale.

In turn, it can be shown that the energy levels appearing in eq.~(\ref{wholeng}) can be treated in a perturbative expansion in powers of $(\sigma R^2)^{-1}$, and the corresponding truncated approximations for $Z(R,L)$ correctly reproduce
 eq.~(\ref{zlo}) at LO (see also \cite{lw04}) and eq.~(\ref{znlo}) at NLO
 (calculations are reported in the Appendix).  
 
\vskip 0.2cm

It is 
particularly
interesting to study the domain of convergence of the 
series 
in (\ref{wholeng}).

For large $n$, the asymptotic behaviour of $P(n)$ is \cite{hardyramanujan}:
\eq
\label{aymptoticpn}
P(n) \sim \frac{1}{4 n \sqrt{3}} e^{\pi \sqrt{ \frac{2n}{3} } } \;\;.
\en
Thus, the series on the r.h.s. of eq.~(\ref{wholeng}) turns out to be convergent for:
\eq
\label{convcorr1}
\frac{1}{\sigma L^2} < \frac{3}{\pi} \;\;.
\en

On the other hand, a convergence constraint over $R$ can be worked out exploiting the open-closed string duality to rewrite $Z(R,L)$ as a series of Bessel functions~\cite{lw04}:
\eq
\label{corr2}
\smt P(n) e^{-E_n(R)L} = \sqrt{\frac{\sigma L^2}{\pi} } \smt P(n) K_0 ( \tilde{E}_n R )
\en
involving $\tilde{E}_n=\tilde{E}_n (L) $, the closed string energies:
\eq
\label{constrainedclosedstringenergies}
\tilde{E}_n (L) = \sigma L \sqrt{ 1+ \frac{8\pi}{\sigma L^2} \left( n - \frac{1}{24} \right)}  \;\;.
\en
Since the asymptotic behaviour of $K_0(x)$ for large values of $x$ is:
\eq
\label{aymptotick0}
K_0(x) \sim \sqrt{\frac{\pi}{2 x}} e^{-x} \left[ 1 + O \left( \frac{1}{x} \right) \right]
\en
the series on the r.h.s. of (\ref{corr2}) converges for:
\eq
\label{convcorr2}
\frac{1}{\sigma R^2} < \frac{12}{\pi} \;\;.
\en

Let us comment on the constraints (\ref{convcorr1}) and (\ref{convcorr2}):
\begin{itemize}
\item They 
can be obtained from each other via interchanging $2R$ and $L$: this ``symmetry'' is related to the open-closed string duality. 
\item They can also be derived by requiring that the 
ground state energies for the open and closed string energies are real; in particular,
inequality (\ref{convcorr1}) --- which was derived requiring convergence for the series involving \emph{open} string states --- can be obtained imposing that the \emph{closed} string ground state energy is real, while inequality (\ref{convcorr2}) --- which, \emph{vice versa}, was obtained requiring convergence of the series involving the \emph{closed} string spectrum --- can be obtained imposing that the \emph{open} string ground state is real.
\item
These bounds 
are the 
analogue of the Hagedorn temperature in string thermodynamics. 
\item
The bound of eq.(\ref{convcorr1}) can be interpreted as a prediction 
for the deconfinement 
temperature\cite{olesen_dec,pa82} leading to the estimate 
$T_c=\sqrt{\frac{3\sigma}{\pi}}$, which is in rather good 
agreement with the results of 
numerical 
simulations for various lattice gauge theories in $d=2+1$. 

\item
The inequality (\ref{convcorr2}) implies a lower bound for the interquark
distance at which the Nambu-Goto string could be observed: 
${\sigma R_c^2} > \frac{\pi}{12}\sim 0.262...$ .
In physical units\footnote{Note that the length scale refers to QCD in 3+1 dimensions.}
$R_c$ is slightly below 0.1 fm (as it can be easily seen
comparing it with
the so called 
``Sommer scale'' $R_s$ defined by the 
  relation $F(R_s) R_s^2 = 1.65$ \cite{Sommer:1993ce}
which corresponds to a physical length $R_c \simeq 0.5$ fm).
However this bound is less relevant from a physical point of view
  than the previous one since, as
 we shall see below, the MC results for the interquark potential start
to diverge from the
predictions of the Nambu-Goto action at distances much larger than $R_c$.
\end{itemize}

\section{Comparison with 
Monte Carlo 
simulations}
\label{comparisonsect}

Before presenting the comparison of theoretical predictions with the numerical results for various gauge models, let us 
discuss a few aspects about the evaluation of the partition function appearing in eq.~(\ref{partitionfunction}) by means of eq.~(\ref{wholeng}).

The sum of the series in eq.~(\ref{wholeng})
cannot be evaluated in closed form
and its numerical estimate is not completely trivial, since 
both rounding and truncation effects must be kept under control.

Calculating the partitions of the integers $P(n)$ appearing as coefficients in
eq.~(\ref{wholeng}) is a task that can be addressed in an efficient way
without resorting to any commercial program
by means of the following identity:
\eq
\label{pnandpkn}
P(n)=\sum_{k=1}^n P_k(n)
\en
where $P_k(n)$ denotes the number of partitions of $n$ made exactly of $k$ parts, which
enjoys the recursive relation~\cite{stanley}:
\eq
\label{pknrecursion}
P_k(n)=P_{k-1}(n-1)+P_k(n-k) \;\;.
\en

To avoid truncation errors one can use from a given threshold $n_{max}$ the
approximation of eq.(\ref{aymptoticpn}) for the weights $P(n)$ and then
approximate the
sum for $n>n_{max}$ in the partition function with an integral. 
We checked that for all the
choices of $\sigma$, $R$ and $L$ that we studied $n_{max}=10000$ was enough to
keep the  systematic errors of our estimates several orders of magnitude
smaller than the statistical uncertainties of the MC estimates of the same
quantities.

\subsection{$\Z_2$ gauge model}

\begin{figure}
\centering
\includegraphics[height=12cm]{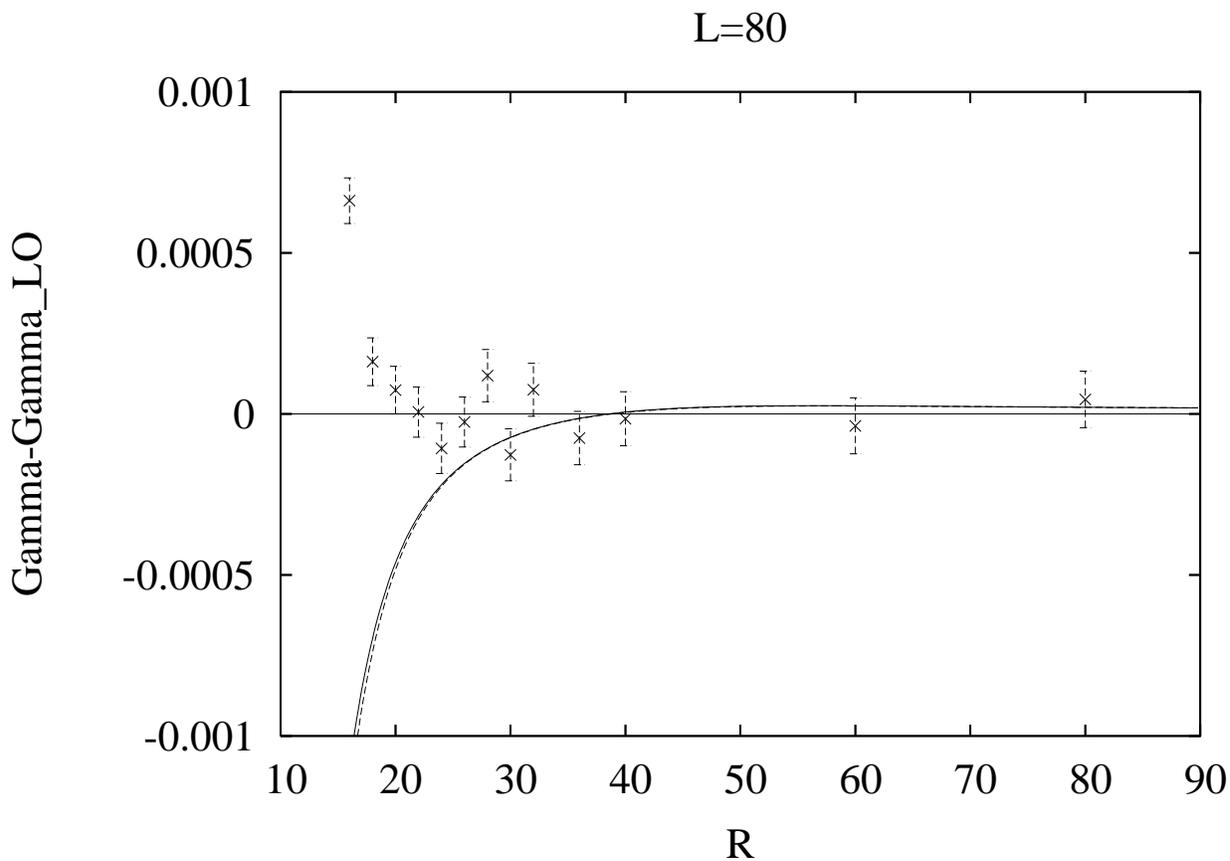}
\caption{Differences  between 
the
values of $\Gamma(R)$ for the full 
Nambu-Goto 
action (dashed line), the NLO approximation 
(solid 
line) and the 
Monte Carlo 
results (crosses) with respect to the LO approximation for the sample at 
$\beta=0.75180$, $L=80$ in the Ising 
gauge
model (data of table~\ref{tab1}).
Notice that for this value of $L$ the NLO and full 
Nambu-Goto
results almost coincide and 
cannot be separated
in the figure.}
\label{fig1}
\end{figure}

\begin{figure}
\centering
\includegraphics[height=12cm]{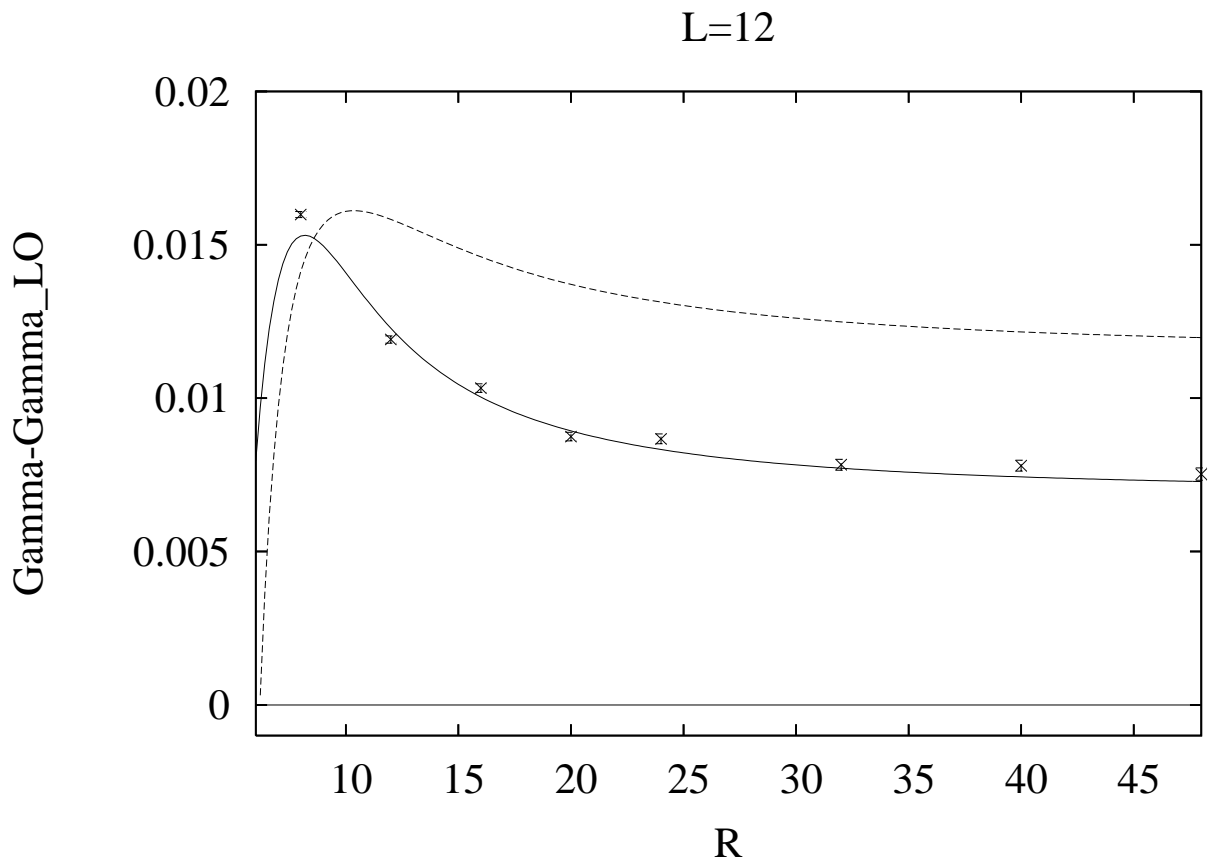}
\caption{Same as figure~\ref{fig1} but for the data at $L=12$ (see table~\ref{tab4}). In this case the 
difference between 
NLO and full 
Nambu-Goto 
predictions 
is 
perfectly detectable.}
\label{fig2}
\end{figure}

As a first test,  
we considered the results for 
$\Z_2$ gauge model in $d=2+1$  
taken from~\cite{chp03} and~\cite{chp04} --- to 
which we refer the reader for notations and general properties of this lattice gauge theory.
In addition to these data we also performed some new simulations whose results are
collected in 
table~\ref{tabrfix}.
In  
the present analysis, we 
focus the attention 
onto
the 
data 
samples 
corresponding to the coupling parameter $\beta=0.75180$ (which is the 
nearest
to the critical
point, among the values studied in
the mentioned papers), for which the finite temperature deconfinement transition occurs 
at $L_c=8$~\cite{ch96}. For the samples that we considered here,
the temperatures range from $T=\frac{T_c}{10}$ (see figure~\ref{fig1}) to $T=\frac{2}{3}T_c$ (see figure~\ref{fig2}).
For the zero temperature string tension at this value of $\beta$,
we used the estimate given 
in ~\cite{chp04}: $\sigma=0.0105241(15)$ which was obtained from our data sample 
at $T=\frac{T_c}{10}$, i.e. $L=80$ and $R\ge 22$, taking into account LO 
order corrections only. 
The numbers given in table \ref{tab1} show that subleading
corrections are of a similar size as the statistical errors of our Monte Carlo
data for these large values of $L$ and $R$.

In our analysis, we
did not consider effects induced by a possible ``boundary term'' that could be included in the effective string action \cite{lw02}, since numerical results in~\cite{chp04} and \cite{cpr04} already showed that the coefficient of such a term is vanishing for this model, and, from the theoretical point of view, the presence of a boundary term in the string action is ruled out, if open-closed string duality holds~\cite{lw04}.

For this value of $\beta$ we expect very small scaling corrections to our data. In~\cite{chp04} we directly
verified this by comparing our results for $\beta=0.75180$ with those taken at $\beta=0.73107$ 
(corresponding to
a string tension which 
is
roughly four times 
larger than here). 
An independent check is given by the
product $\sigma L_c^2$.  For the value of $\beta$ 
we study 
here,
one finds $\sigma L_c^2=0.6740(10)$,
which deviates from the asymptotic estimate 
$\sigma L_c^2=0.654(3)$ less than $3\%$~\cite{Martin_thesis}.

The results of our analysis are 
shown in tables~\ref{tab1}--\ref{tabrfix}, where
Monte Carlo results for the ratio $\Gamma(R)=\frac{G(R+1)}{G(R)}$ (rightmost column)
are compared with the corresponding predictions from the LO-truncated, NLO-truncated, or full effective Nambu-Goto string, obtained from eq.~(\ref{zlo}), from eq.~(\ref{znlo}), and from eq.~(\ref{wholeng}), respectively.
 
All these theoretical estimates are affected by an
uncertainty (due to the statistical error in $\sigma$, and proportional to $\sigma L^2$)
whose relative amount is of order $10^{-5}$ (for $L=12$) or $10^{-4}$ (for $L=80$), and it is never larger than the statistical errors of the
Monte Carlo data. For this reason, we chose not to report this overall uncertainty in the tables, in order to avoid confusion.

We remark the fact 
that 
the comparison is ``absolute'', in the sense that there is no free parameter.

\begin{figure}
\centering
\includegraphics[height=12cm]{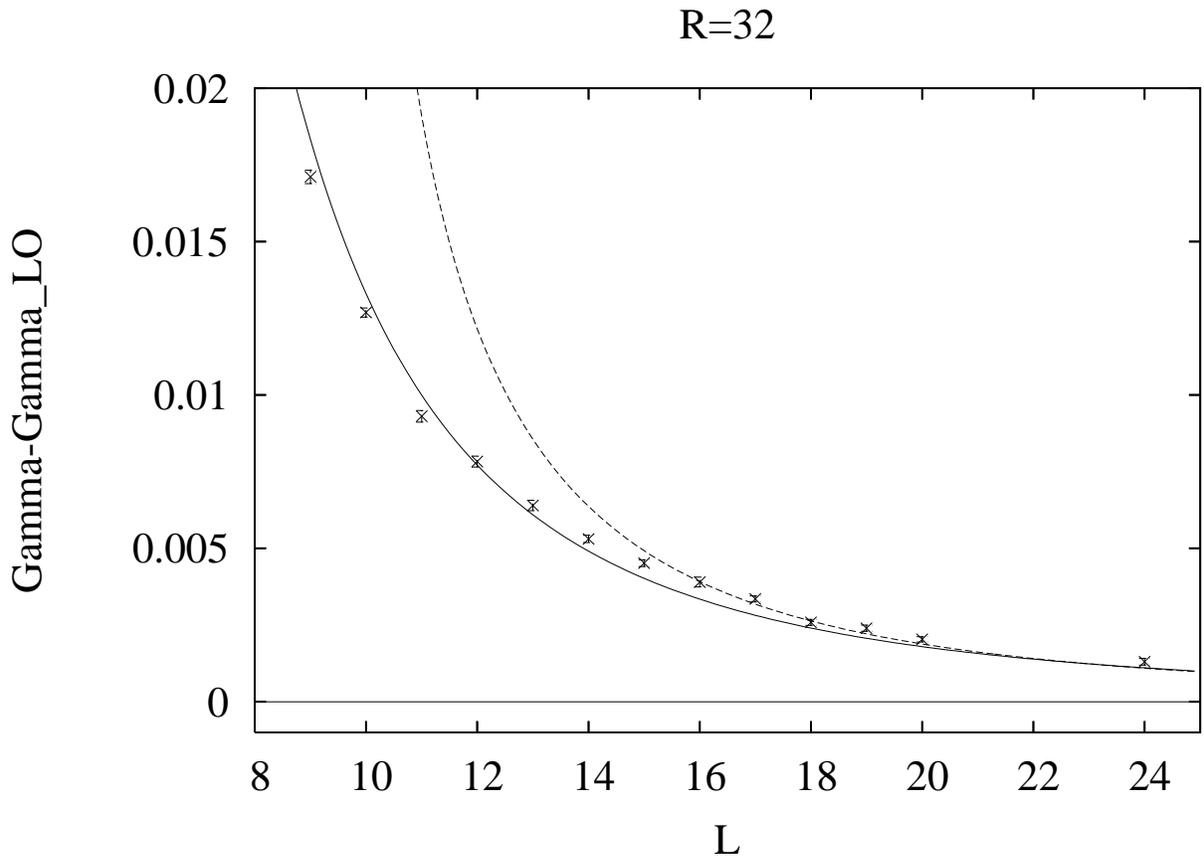}
\caption{Same as figure~\ref{fig1}, but for the data at $R=32$ (see table~\ref{tabrfix}). As mentioned in the text, in this case the 
Monte Carlo 
results 
appear to interpolate between
the full 
Nambu-Goto
behaviour (dashed line) at low temperature
and
the NLO one
(solid 
line) at high temperature.}
\label{fig3}
\end{figure}

The 
results can be summarized as  follows.
\begin{itemize}
\item
For the two samples at the lowest temperatures ($T=\frac{T_c}{10}$ and $T=\frac{T_c}{3}$),
Monte Carlo data and theoretical
estimates
obtained from the full
partition function are in good agreement for $R \geq 30 $ and $R \geq 24$, respectively.
However, it is also important to notice that in this range the NLO-truncated predictions show no substantial difference with respect to the estimates from the complete action.
Conversely, for lower values of $R$, the 
numerical data 
are not compatible with predictions from either eq.~(\ref{zlo}), eq.~(\ref{znlo}), or from eq.~(\ref{wholeng}): in that regime, the Monte Carlo results for $\Gamma (R)$ appear to be systematically larger than the values predicted by the effective string picture,
and, 
in particular, the discrepancy 
gets even larger when $R$ decreases. This 
mismatch
had already been  
observed in~\cite{chp04}, using the NLO-truncation. Here, we find that keeping into account the   
subleading corrections from the whole 
Nambu-Goto action does not remove 
the disagreement\footnote{Quantitatively, the subleading corrections induce an even larger mismatch.}.
\item 
For the two samples at higher temperatures ($T=\frac{T_c}{2}$ 
and $T=\frac{2}{3}T_c$), the contribution  
by subleading terms beyond the NLO becomes quantitatively more relevant
and the precision of the numerical data allows to appreciate 
the difference between the 
predictions of eq.~(\ref{znlo}) and eq.~(\ref{wholeng}) 
for all values of $R$ that are considered. 

For $T=\frac{T_c}{2}$ and $R\ge16$, the Monte Carlo data lie in between 
the full Nambu-Goto expectation and the NLO truncation. 
For $T=\frac{T_c}{2}$  and $R\ge16$, the Monte Carlo data fall
(incidentally), within 
about two standard deviations, on top of the NLO truncation. The numerical
matching with the full Nambu-Goto is clearly worse.

\item
It is very interesting to look at the data at fixed $R=32$ collected in 
table~\ref{tabrfix} (see figure~\ref{fig3}). We see that for all
values of $L>16$ the data agree with the 
Nambu-Goto 
predictions almost within one standard deviation. Fortunately
in this region the precision is high enough to distinguish between LO, NLO and full 
Nambu-Goto. The picture which
emerges, in agreement with the above observation
 is that the 
Nambu-Goto 
string is the correct description for low enough temperatures and that around 
$T=T_c/2$ there is a smooth crossover towards a behaviour which is instead very near to the NLO result. 
We shall discuss in more detail
this point in the last section.

\end{itemize}

\begin{table}[h]
\begin{center}
\begin{tabular}{|r|l|l|l|l|}
\hline
\multicolumn{1}{|c}{$R$}
&\multicolumn{1}{|c}{LO}
&\multicolumn{1}{|c}{NLO}
&\multicolumn{1}{|c}{full NG}
&\multicolumn{1}{|c|}{MC data} \\
\hline
 8 &  0.372554    &    0.358712 &     0.353033 & 0.382613(58)  \\
10  & 0.391751    &    0.385458 &     0.383979 &0.396313(61)  \\
12  & 0.402904      &  0.399676  &    0.399177 &0.405088(65)  \\
14  & 0.409919      &  0.408105  &    0.407905 &0.411017(69)  \\
16  & 0.414606     &   0.413511 &     0.413422 &0.415268(71)  \\
18  & 0.417886      &  0.417189 &     0.417146 &0.418048(74)  \\
20  & 0.420270       & 0.419808  &    0.419786 &0.420344(74)  \\
\hline
\hline
22  & 0.422056      &  0.421741 &     0.421730 &0.422062(78)  \\
24  & 0.423430      &  0.423211 &     0.423205 &0.423323(78)  \\
26  & 0.424511      &  0.424357 &     0.424354  &0.424486(78)  \\
28  & 0.425377       & 0.425270 &     0.425269 &0.425496(81)  \\
30  & 0.426085      &  0.426012 &     0.426011  &0.425958(81)  \\
32  & 0.426672      &  0.426624 &     0.426624  &0.426747(82)  \\
36  & 0.427588      &  0.427574 &     0.427573  &0.427513(83)  \\
40  & 0.428270      &  0.428276 &     0.428274 &0.428255(84)  \\
60  & 0.430134      &  0.430159  &    0.430158 &0.430097(87)  \\
80  & 0.431023     &   0.431043 &     0.431043 &0.431068(88)  \\
\hline
\end{tabular}
\end{center}
\caption 
{\sl Comparison of Monte Carlo results (last column) for correlator ratios $\Gamma(R)=\frac{G(R+1)}{G(R)}$ in the $\Z_2$ gauge model at $\beta=0.75180$ and $T=\frac{T_c}{10}$ (corresponding to $L=80$) with the   
LO-truncated, NLO-truncated, and full predictions from the Nambu-Goto effective string action (respectively: second, third, and fourth column), as a function of the interquark distance in lattice units $R$ (first column).
Note that the data for $R\ge 22$ were used in ref. \cite{chp04}
to determine the value of $\sigma$ which is used here; assuming LO corrections.}
\label{tab1}
\end{table}

\begin{table}[h]
\begin{center}
\begin{tabular}{|r|l|l|l|l|}
\hline
\multicolumn{1}{|c}{$R$}
&\multicolumn{1}{|c}{LO}
&\multicolumn{1}{|c}{NLO}
&\multicolumn{1}{|c}{full NG}
&\multicolumn{1}{|c|}{MC data} \\
\hline
  8  &  0.74374 &  0.73654 &   0.73364 & 0.75075(7) \\
 12  &  0.76220 &  0.76282 &   0.76223 & 0.76492(8) \\
 16  &  0.77015 &  0.77183 &   0.77150 & 0.77193(8) \\
 20  &  0.77479 &  0.77636 &   0.77625 & 0.77661(9) \\
 24  &  0.77789 &  0.77925 &   0.77925 & 0.77926(10) \\
 32  &  0.78181 &  0.78291 &   0.78297 & 0.78311(11) \\
 40  &  0.78419 &  0.78516 &   0.78523 & 0.78532(12) \\
 48  &  0.78578 &  0.78669 &   0.78677 & 0.78677(13) \\
\hline
\end{tabular}
\end{center}
\caption 
{\sl Same as in table \ref{tab1}, but for $T=\frac{T_c}{3}$ ($L=24$).}
\label{tab2}
\end{table}

\begin{table}[h]
\begin{center}
\begin{tabular}{|r|l|l|l|l|}
\hline
\multicolumn{1}{|c}{$R$}
&\multicolumn{1}{|c}{LO}
&\multicolumn{1}{|c}{NLO}
&\multicolumn{1}{|c}{full NG}
&\multicolumn{1}{|c|}{MC data} \\
\hline
  8  & 0.82236 & 0.82516 & 0.82161  & 0.83207(8) \\
 12  & 0.83885 & 0.84492 & 0.84415  & 0.84523(9) \\
 16  & 0.84707 & 0.85205 & 0.85233  & 0.85220(10) \\
 20  & 0.85210 & 0.85632 & 0.85693  & 0.85655(12) \\
 24  & 0.85550 & 0.85928 & 0.85999  & 0.85937(13) \\
 32  & 0.85981 & 0.86315 & 0.86390  & 0.86371(16) \\
 40  & 0.86242 & 0.86556 & 0.86631  & 0.86624(15) \\
 48  & 0.86418 & 0.86721 & 0.86796  & 0.86740(17) \\
\hline
\end{tabular}
\end{center}
\caption 
{\sl Same as in table \ref{tab1}, but for $T=\frac{T_c}{2}$ ($L=16$).}
\label{tab3}
\end{table}

\begin{table}[h]
\begin{center}
\begin{tabular}{|r|l|l|l|l|}
\hline
\multicolumn{1}{|c}{$R$}
&\multicolumn{1}{|c}{LO}
&\multicolumn{1}{|c}{NLO}
&\multicolumn{1}{|c}{full NG}
&\multicolumn{1}{|c|}{MC data} \\
\hline
  8  &  0.86788     &   0.88317  &  0.88202 & 0.88387(10)  \\
 12  &  0.88455     &   0.89683  &  0.90037 & 0.89646(12) \\
 16  &  0.89318     &   0.90322  &  0.90779 & 0.90351(14) \\
 20  &  0.89848     &   0.90741  &  0.91219 & 0.90723(14) \\
 24  &  0.90207     &   0.91039  &  0.91520 & 0.91074(16) \\
 32  &  0.90661     &   0.91433  &  0.91910 & 0.91444(18) \\
 40  &  0.90937     &   0.91680  &  0.92153 & 0.91717(18) \\
 48  &  0.91122     &   0.91851  &  0.92320 & 0.91875(20) \\
\hline
\end{tabular}
\end{center}
\caption 
{\sl Same as table \ref{tab1}, but for $T=\frac{2}{3}T_c$ (corresponding to $L=12$).}
\label{tab4}
\end{table}

\begin{table}[h]
\begin{center}
\begin{tabular}{|r|l|l|l|l|}
\hline
\multicolumn{1}{|c}{$L$}
&\multicolumn{1}{|c}{1 loop }
&\multicolumn{1}{|c}{2 loop}
&\multicolumn{1}{|c}{full NG}
&\multicolumn{1}{|c|}{MC data} \\
\hline
 9 &   0.94940     &      0.96772 &          & 0.96651(22) \\
10  &  0.93401    &     0.94732 &  0.98210   & 0.94670(15)\\
11  &  0.91984    &     0.92984 &  0.93933   & 0.92915(19)\\
12  &  0.90661   &     0.91433 &   0.91910  & 0.91444(18) \\
13   & 0.89411    &    0.90021 &   0.90295   & 0.90051(17)  \\
14  &  0.88221    &    0.88712 &   0.88882  &  0.88752(12)  \\
15   & 0.87080  &    0.87482 &     0.87593  & 0.87532(11)  \\
16   & 0.85981     &    0.86315 &  0.86390   &  0.86371(16)  \\
17   & 0.84917    &    0.85199 &   0.85251   & 0.85252(10)   \\
18   & 0.83884    &    0.84124 &   0.84161  & 0.84143(10)   \\
19   & 0.82879   &    0.83086 &    0.83113 & 0.83119(10)\\
20   & 0.81899    &    0.82078 &   0.82098  & 0.82102(09)\\
24  &  0.78181     &    0.78291 &  0.78297  & 0.78311(11) \\

\hline
\end{tabular}
\end{center}
\caption 
{\sl Same as table \ref{tab1}, but for a fixed value of $R$: $R=32$ and various values of $L$.}
\label{tabrfix}
\end{table}

\subsection{$SU(2)$ gauge model}

For the $SU(2)$ model  in $d=2+1$ dimensions we considered data published in~\cite{cpr04} 
(to which we refer the reader for details and notation), 
focusing onto the two samples corresponding to the largest ($L=60$) and the smallest ($L=8$) 
values the inverse temperature at $\beta=9.0$ --- the other samples reported in~\cite{cpr04} 
show the same qualitative behaviour which is observed for the data set at  $L=60$, 
adding no further information to the present analysis.

For $\beta=9.0$ the string tension of this model is known to a high degree of precision: 
$\sigma=0.025900(12)$ \cite{cpr04}. The lower bound $R_c$ is between 3 and 4 lattice spacings
and the deconfinement temperature in lattice units is approximately $T_c \simeq \frac{1}{6}$, thus the two samples that we are considering correspond to $T=\frac{T_c}{10}$ and $T=\frac{3}{4}T_c$ respectively. In agreement with
notations in~\cite{cpr04}, we define the following quantity:
 \eq
 Q(R)= -\frac{1}{L} \log\left(\frac{G(R+1)}{G(R)}\right)
\en
whose dominant contribution is the string tension; the relative error induced by the uncertainty in 
$\sigma$ is approximately of order 
$10^{-4}$, which turns out to be negligible or comparable with respect to the Monte Carlo result precision for both 
samples. 

Comparison between numerical data and theoretical expectations is reported in tables~\ref{tab5},~\ref{tab6}, 
with the same format used for previous tables.

\begin{figure}
\centering
\includegraphics[height=12cm]{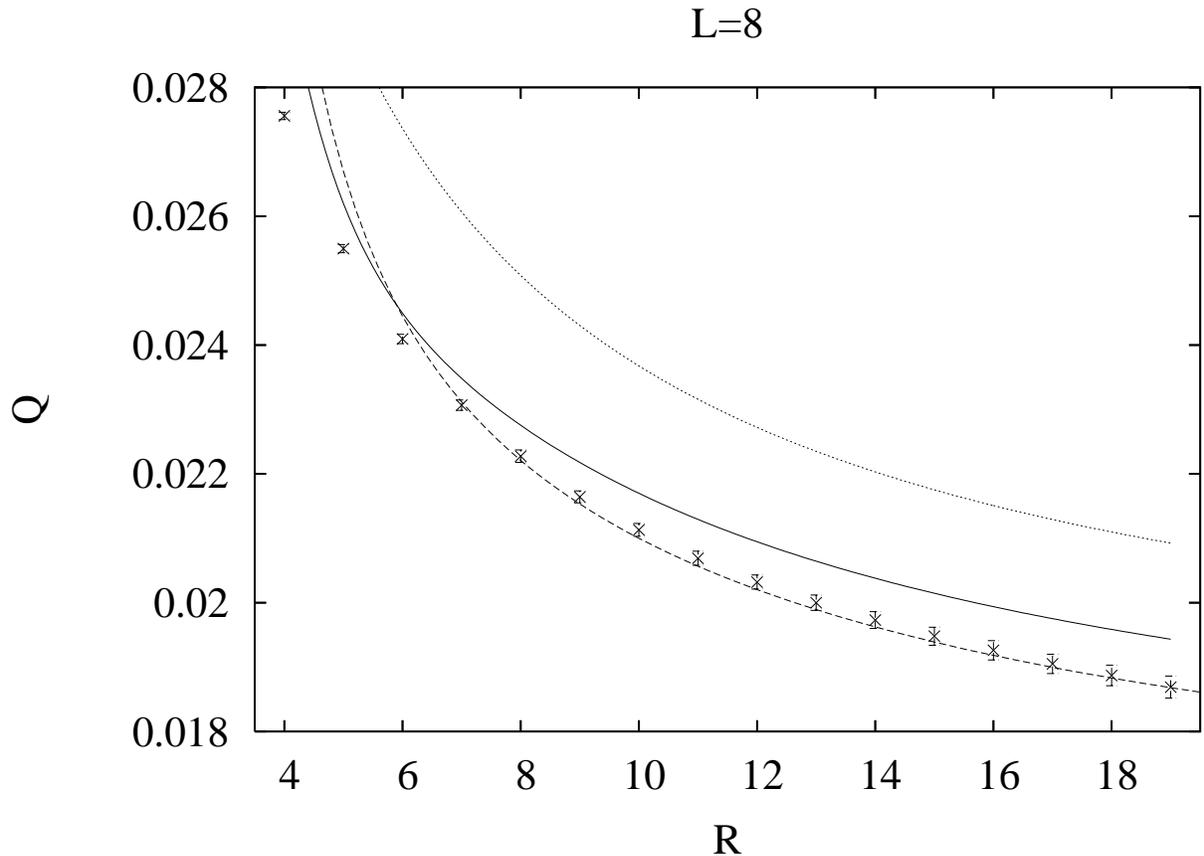}
\caption{Values of $Q(R)$ for the full 
Nambu-Goto
action (dashed line), the NLO approximation
(solid 
line), the LO approximation (dotted line) and the
Monte Carlo
results (crosses) for the sample at $\beta=9.0$, $L=8$ in the 
$SU(2)$
model (data of table~\ref{tab6}).}
\label{fig4}
\end{figure}

The results for this model show that:
\begin{itemize}
\item
At low temperatures, the behaviour of Monte Carlo data for $SU(2)$ is very similar to the 
$\Z_2$ gauge model; in
particular, the lattice simulation results disagree with respect to the
predictions from the full Nambu-Goto partition function, 
whose subleading terms with respect to the NLO-truncated approximation 
induce an increasing
gap between numerical data and theoretical expectations 
as shorter and shorter interquark 
distances are investigated.
 \item
On the contrary, for the high temperature ($T=\frac{3}{4}T_c$) sample, $SU(2)$ results are in remarkable agreement with the predictions of the full Nambu-Goto partition function: agreement within statistical errors is observed for all interquark distances $R>6$ (see
figure~\ref{fig4}). Notice that in this temperature regime the subleading string effects play a relevant role, and the LO, NLO and full Nambu-Goto predictions can be clearly distinguished within the precision of numerical data: in particular, the difference between the full-order prediction and the NLO approximation is larger than five times the uncertainty of Monte Carlo data, for all values of $R$.

\end{itemize}

\begin{table}[h]
\begin{center}
\begin{tabular}{|r|l|l|l|l|}
\hline
\multicolumn{1}{|c}{$R$}
&\multicolumn{1}{|c}{LO}
&\multicolumn{1}{|c}{NLO}
&\multicolumn{1}{|c}{full NG}
&\multicolumn{1}{|c|}{MC data} \\
\hline
  2 &  0.047717    &      0.076813 &            &0.040455(7)\\
  3 &  0.036808     &     0.043891 &             &0.034374(10)\\
  4 &  0.032445     &     0.034967  & 0.037080   &0.031426(12)\\
  5 &  0.030263    &      0.031378  & 0.031842  &0.029785(14)\\
  6 &  0.029017    &      0.029584  & 0.029731  &0.028777(16)\\
  7 &  0.028237   &       0.028556  & 0.028614  &0.028113(17)\\
  8 &  0.027718     &     0.027910  & 0.027936 &0.027654(19)\\
  9 &  0.027354      &    0.027477  & 0.027490 &0.027322(22)\\
 10 &  0.027090     &     0.027172  & 0.027179 &0.027076(24)\\
 11 &  0.026892      &    0.026949  & 0.026953 &0.026887(28)\\
 12 &  0.026739     &     0.026780  & 0.026782 &0.026739(33)\\
 13 &  0.026619    &      0.026649  & 0.026651 &0.026616(40)\\

\hline
\end{tabular}
\end{center}
\caption 
{\sl 
Monte Carlo results (rightmost column) for $Q(R)$ 
in $SU(2)$ gauge theory  
at $\beta=9.0$ and $T=\frac{T_c}{10}$ (corresponding to $L=60$)
in comparison with the theoretical expectations from the LO (second column), NLO (third column) and full (fourth column) effective Nambu-Goto action.
}
\label{tab5}
\end{table}

\begin{table}[h]
\begin{center}
\begin{tabular}{|r|l|l|l|l|}
\hline
\multicolumn{1}{|c}{$R$}
&\multicolumn{1}{|c}{LO}
&\multicolumn{1}{|c}{NLO}
&\multicolumn{1}{|c}{full NG}
&\multicolumn{1}{|c|}{MC data} \\
\hline
2&   0.04769   &     0.07537 &            &    0.03741(4) \\
3&   0.03660  &      0.04001 &            &    0.03093(5)\\
4&   0.03185  &      0.02992  &   0.03179 &    0.02756(6)\\
5&   0.02915    &    0.02620 &    0.02670 &    0.02550(6)\\
6&   0.02736   &     0.02449  &   0.02445 &    0.02409(7)\\
7&   0.02607   &     0.02348 &    0.02312 &    0.02307(8)\\
8&   0.02508   &     0.02275 &    0.02221 &    0.02227(9)\\
9&   0.02430    &    0.02218  &   0.02153 &    0.02164(9)\\
10&  0.02368   &     0.02170 &    0.02100 &    0.02113(10)\\
11&  0.02316   &     0.02129  &   0.02057 &    0.02069(11)\\
12&  0.02272   &     0.02095  &   0.02020 &    0.02032(11)\\
13&  0.02235   &     0.02064  &   0.01990 &    0.02000(12)\\
14&  0.02203  &      0.02038 &    0.01962 &    0.01973(13)\\
15&  0.02175   &     0.02015  &   0.01939 &    0.01948(14)\\
16&  0.02151  &      0.01994  &   0.01918 &    0.01926(15)\\
17&  0.02129   &     0.01975 &    0.01900 &    0.01905(15)\\
18&  0.02110  &      0.01958 &    0.01883 &    0.01887(16)\\
19&  0.02092  &      0.01943 &    0.01868 &    0.01869(17)\\

\hline
\end{tabular}
\end{center}
\caption 
{\sl Same as in table \ref{tab5}, but for $T=\frac{3}{4}T_c$ ($L=8$).}
\label{tab6}
\end{table}

\subsection{$SU(3)$ gauge model}
For the $SU(3)$ model  in $d=2+1$ dimensions,
we compare the theoretical expectations with the data published in~\cite{lw02}. 
We focused onto the sample at $\beta=20$, using the data reported in table 3 of \cite{lw02} to build the $Q(R)$ observable discussed above\footnote{Notice the shift in $R$ between our $Q(R)$ and the quantity $F(R)$ in~\cite{lw02}.}. In this case we have $L=60$, corresponding to a temperature $T\simeq \frac{T_c}{10}$, while we have no data for the high $T$ regime. An estimate for the
string tension $\sigma$ can be obtained from the data themselves, following a procedure similar to the one used in~\cite{cpr04} for the $SU(2)$ model; 
we found $\sigma=0.033754(3)$,
corresponding to 
a lower bound $R_c$ between 2 and 3 lattice spacings.

Comparison of theoretical predictions and numerical data is reported in table \ref{tab7},
in the same format used for the previous cases. The pattern which emerges is the same that is observed for $SU(2)$ at low temperatures, the only difference
being that the mismatch between the full Nambu-Goto string prediction and Monte Carlo results appears to be slightly smaller than
in the $SU(2)$ case. 

\begin{table}[h]
\begin{center}
\begin{tabular}{|r|l|l|l|l|}
\hline
\multicolumn{1}{|c}{$R$}
&\multicolumn{1}{|c}{LO}
&\multicolumn{1}{|c}{NLO}
&\multicolumn{1}{|c}{full NG}
&\multicolumn{1}{|c|}{MC data} \\
\hline
   2  &  0.0555706     &    0.0778972    &   & 0.0508627(32)  \\
   3  &  0.0446623    &     0.0500971 &  0.0592690  & 0.0434942(32)  \\
   4  &  0.0402990     &    0.0422344 &  0.0432510  & 0.0399646(33)   \\
   5  &  0.0381173    &     0.0389728 &  0.0392195  & 0.0380274(35)    \\
   6  &  0.0368707     &    0.0373058 &  0.0373874  & 0.0368546(38)   \\
   7  &  0.0360915   &      0.0363358&   0.0363683  & 0.0360941(41)   \\
   8  &  0.0355721    &     0.0357196 &  0.0357344  & 0.0355749(45)    \\
   9  &  0.0352084    &     0.0353027 &  0.0353102  & 0.0352080(49)   \\
  10  &  0.0349440      &   0.0350071  & 0.0350111  & 0.0349401(55)   \\
  11  &  0.0347457     &    0.0347895  & 0.0347917  & 0.0347382(62)  \\
  12  &  0.0345931      &   0.0346244  & 0.0346258  & 0.0345838(73)  \\
  13  &  0.0344732     &    0.0344962 &  0.0344970  & 0.0344673(92)   \\
\hline
\end{tabular}
\end{center}
\caption 
{\sl Comparison between theoretical predictions and numerical results for the $SU(3)$ model in $d=2+1$ dimensions obtained from data published in \cite{lw02}. Results refer to $\beta=20.0$, $L=60$ (corresponding to $T \simeq \frac{T_c}{10}$). The format is the same as for table \ref{tab5} .}
\label{tab7}
\end{table}

\section{Conclusions}
\label{conclusionsect}

The pattern emerging from the comparisons discussed in the previous section can be summarized in the following points.
\begin{itemize}
\item
The Nambu-Goto action describes Monte Carlo data rather well for large values of $L$ and $R$; however, in this regime it is almost indistinguishable from its truncation at NLO and (for very large values of $R$) from the pure free bosonic effective string.
Nevertheless, it is interesting to note that results from our simulations of the Ising gauge theory at fixed $R$ ($R=32$ in our case) show that the predictions of the full Nambu-Goto action match the numerical data in the range $L \ge 16$, and in particular there exists a window ($16 \le L \le 20$) where the full-order and the NLO-truncated predictions can be clearly distinguished --- while for larger values of $L$ this is not possible anymore, within our statistical precision.
\item
At low temperatures, a common pattern for the three models is observed as $R$ decreases: 
clear evidence of a mismatch with respect to the string expectation is found. 
Such a mismatch is increasingly severe for shorter and shorter interquark distances
and is not fixed by keeping into account subleading corrections beyond the NLO approximation 
of the Nambu-Goto string action.
This observation has two relevant 
consequences. 
First, 
at short distances the string spectrum must be
different from the Nambu-Goto one, in agreement with the recent observations of~\cite{Juge:2002br,jkm03}. 
Furthermore,
the short distance behaviour seems to disagree with
 the results obtained in~\cite{lw04} by imposing the 
open-closed string duality 
of
the effective string. In fact 
in~\cite{lw04} it was shown that in $d=2+1$ the NLO contribution to the energy levels is fixed and must coincide with the 
Nambu-Goto one. This implies (see the derivation in the Appendix) 
  that also the NLO contribution to the interquark potential must be the 
Nambu-Goto one for any effective string theory fulfilling open-closed string duality.  A possible interpretation of these two observations is that at these scales (despite the fact that the leading order L\"uscher term still correctly describes the gross
  features of the interquark potential) there is a breakdown of the effective 
  string picture~\cite{Juge:2002br,jkm03}.

\item
For lower values of $L$ (i.e. as the deconfinement temperature is approached) the situation 
seems to change, and the effective string picture appears to provide a better description of Monte Carlo data as $T$ approaches $T_c$. However, there is 
a relevant difference between the behaviour observed in the $\Z_2$ and in the $SU(2)$ gauge models: for the Ising gauge model, data are in better agreement with the NLO-truncated than with the full partition function predictions, whereas for the $SU(2)$ theory, the whole Nambu-Goto action appears to provide a remarkably good description of the numerical results. 
\item
However, it is important to stress that this agreement could well be only a coincidence. In fact, dimensional
reduction and the Svetitsky-Yaffe conjecture~\cite{sy82} suggest a critical index for the deconfinement
transition (both for the Ising and for the $SU(2)$ model) equal to the one of the 2d spin Ising model, i.e.
$\nu=1$. This means:
\eq
\sigma(T)\sim \sigma_0(T-T_c)
\en
On the contrary,
 from the 
Nambu-Goto action we find $\nu=1/2$~\cite{olesen_dec,pa82}. This suggests that at some stage there should be
a truncation of the square root expansion in the Nambu-Goto effective string. The order of the power at which
this truncation occurs needs not to be universal and can be inferred by looking at the critical temperature. In
the Nambu-Goto model in $d=2+1$ dimensions,  
$T_c$ is expected to be~\cite{olesen_dec,pa82}:
\eq
\label{tcsigma}
\frac{T_c}{\sqrt{\sigma}}=\sqrt{\frac3\pi} \simeq 0.977
\en
This prediction does not agree with the high precision results obtained 
in~\cite{Martin_thesis}  for the $\Z_2$ gauge 
model $\frac{T_c}{\sqrt{\sigma}}=1.237(3)$. 
If one evaluates the order at which the 
Nambu-Goto expansion
should be truncated in order to obtain this value for the critical temperature one finds with a good
approximation  a quartic 
polynomial, i.e. one expects to describe well the data with the leading and next to
leading order only, as we saw in the previous section. This is particularly evident if one looks at the data
collected in 
table~\ref{tabrfix}\footnote{Notice however that 
it is rather non trivial and in principle unexpected the fact that the linear critical
behaviour expected for the model is exactly realized (from a numerical point of view) truncating the Nambu-Goto
action and not with a completely different coefficient in front of the quartic term.}.
On the contrary in the 
$SU(2)$ case the estimate for $T_c$
is  
closer to the 
value predicted by the Nambu-Goto model. The most accurate estimate is (for a recent review with updated estimates
see~\cite{mt04}) $\frac{T_c}{\sqrt{\sigma}}=1.12(1)$,
corresponding to much a higher order in the truncation
and,
accordingly,
a better agreement with the prediction of the 
all-order 
Nambu-Goto prediction. 

\end{itemize}

The picture which emerges is that the 
Nambu-Goto action is most probably
 the correct effective description for large enough values of $R$
and $L$ (see in particular the data collected in 
table~\ref{tabrfix}). However, for different reasons, as $R$ or
$L$ are decreased this effective description looses its validity. When $R$ is decreased 
(most probably due to the anomaly problem discussed in 
sect.~2) the effective string picture as a whole seems to
be no longer valid. When $L$ is decreased the 
Nambu-Goto string which predicts mean field critical indices must
necessarily break down and we observe a crossover towards a new string behaviour (which is different for
 different gauge groups) with a  critical behaviour compatible with that expected from dimensional reduction and
 the 
Svetitsky-Yaffe 
conjecture. In particular, in 
 the two cases that we studied: 
$\Z_2$ and $SU(2)$, 
this effective string
 seems to be a
 suitable truncation of the whole Nambu-Goto action.
Summarizing these observations we could look at 
the Nambu-Goto action as a sort of ``mean field effective string'' which turns out to be a remarkably good
approximation for distances larger than a few times the bulk correlation length of the model.

In this respect it would be very interesting 
to extend 
the high temperature analysis  also
to the $SU(3)$ gauge model,
running low $L$ simulations 
like to the ones presented in~\cite{cpr04} for the $SU(2)$ gauge group.
In fact in the $SU(3)$ case we have (see~\cite{mt04})
$\frac{T_c}{\sqrt{\sigma}}=0.98(2)$, i.e. an almost perfect agreement of the critical temperature 
with the Nambu-Goto expectation. Moreover in the 
$SU(3)$
case the truncation mechanism suggested above should not
work since in this case we expect a non-integer critical index for the
deconfinement transition (which according to the  
Svetitsky-Yaffe 
conjecture should belong to the 2d 
$\Z_3$ 
universality class).

\vskip1.0cm {\bf
Acknowledgements.} This work was partially supported by the European Commission TMR programme HPRN-CT-2002-00325 (EUCLID). M.~P. acknowledges support received from Enterprise Ireland under the Basic Research Programme.

\newpage
\appendix{}
\label{app}
\vskip 0.5cm
\section{{{Partition function truncated at LO and NLO}}}
\renewcommand{\theequation}{A.\arabic{equation}}
\setcounter{equation}{0}

We show the matching between the full partition function eq.~(\ref{wholeng}) and the LO- and NLO-truncated approximations eq. (\ref{zlo}) and eq. (\ref{znlo}) respectively.

The square root appearing in the open string spectrum eq.~(\ref{arvisspectrum}) can be expanded perturbatively in powers of $(\sigma R^2)^{-1}$:
\eq
\label{perturbativearvisspectrum}
E_n (R) = \sigma R + \frac{\pi}{R} \left( n- \frac{1}{24} \right) - \frac{\pi^2}{2 \sigma R^3} \left( n- \frac{1}{24} \right)^2 + O \left(\sigma^{-2} R^{-5} \right)
\en
Truncating at the first non-trivial order yields:
\eq
\label{arvisspectrumlo}
E^{LO}_n (R) = \sigma R + \frac{\pi}{R} \left( n- \frac{1}{24} \right)
\en
and, correspondingly, the leading-order truncation of the partition function 
eq.~(\ref{wholeng}) reads:
\eq
\label{wholenglo1}
Z^{LO}(R,L)=e^{-\sigma RL} e^{+\frac{\pi}{24}\frac{L}{R}}\sum_{n=0}^\infty P(n)
e^{-\frac{\pi Ln}{R}}
\en
Defining: $q\equiv e^{2\pi i \tau}$, with $\tau=iu=i\frac{L}{2R}$, one finds:
\eq
\label{wholenglo2}
Z^{LO}(R,L)=e^{-\sigma RL} q^{-\frac{1}{24}}\sum_{n=0}^\infty P(n)
q^{n}
\en
Due to the well-known identity:
\eq
\frac{1}{\prod_{n=1}^\infty(1-q^n)}=\sum_{n=0}^\infty P(n) q^{n}
\label{id}
\en
and to Dedekind's function definition eq.~(\ref{eta}), we end up with: 
\eq 
\label{wholenglo3}
Z^{LO}(R,L)=\frac{e^{-\sigma RL}}{\eta(i u)}
\en
in agreement with eq.~(\ref{zlo}).

On the other hand, truncation of eq.~(\ref{perturbativearvisspectrum}) at the next-to-leading order yields:
\eq
\label{arvisspectrumlo}
E^{NLO}_n (R) = \sigma R + \frac{\pi}{R} \left( n- \frac{1}{24} \right) - \frac{\pi^2}{2 \sigma R^3} \left( n- \frac{1}{24} \right)^2 
\en
Correspondingly, the next-to-leading-order approximation for the partition function reads:
\eq 
\label{wholengnlo1}
Z^{NLO}(R,L)=\frac{e^{-\sigma RL}}{\eta(i u)}\left[1+\frac{\pi^2 L}{1152\sigma R^3}\frac{1}{G(q)}
\sum_{n=0}^\infty P(n) q^n(1-48{n}+576 n^2)\right]
\en
where we defined:
\eq
\label{g}
G(q)\equiv \sum_{n=0}^\infty P(n) q^{n}
\en
The non-trivial sums appearing on the r.h.s. of eq.~(\ref{wholengnlo1}) can be written as:
\eq
\frac{1}{G(q)}\sum_{n=1}^\infty nP(n) q^{n}=q\frac{d}{dq} \ln G(q) \label{nqn}
\en
and:
\eq
\frac{1}{G(q)}\sum_{n=1}^\infty n^2 P(n) q^{n}=q^2\frac{d^2}{dq^2} \ln G(q) +
\frac{1}{G(q)}\sum_{n=1}^\infty nP(n) q^n+
\left[\frac{1}{G(q)}\sum_{n=1}^\infty nP(n) q^n\right]^2 \label{n2qn}
\en
On the other hand, by virtue of the identity (\ref{id}):
\eq
\label{qdqg}
q\frac{d}{dq}\ln G(q)=-\sum_{n=1}^\infty
q\frac{d}{dq} \ln (1-q^n)=\sum_{n=1}^\infty
\frac{nq^n}{1-q^n}
\en
For any positive integer $l$, the
following 
identity holds:
\eq
\label{identitysigmal}
\sum_{n=1}^\infty
\frac{n^lq^n}{1-q^n}=\sum_{k=1}^\infty
\sigma_l(k)q^k
\en
where $\sigma_l(k)$ denotes the sum of the $l$-th powers of the positive divisors of $k$;
this identity can be easily proven by expanding the l.h.s of eq.~(\ref{identitysigmal}):
\eq
\sum_{n=1}^\infty
\frac{n^lq^n}{1-q^n}=\sum_{n=1}^\infty n^l\sum_{k=1}^\infty q^{nk} =\sum_{k=1}^\infty \left(\sum_{d|k}d^l\right) q^k =\sum_{k=1}^\infty \sigma_l(k)q^k
\en
where $\sum_{d|k}$ is precisely the sum over positive divisors of $k$.
Using the identity (\ref{identitysigmal}) and the definition of the Eisenstein function eq.~(\ref{eisenstein2}), one gets:
\eq
q\frac{d}{dq} \ln G(q)=\frac{1-E_2(q)}{24}
\label{use2}
\en
Finally, the terms in (\ref{n2qn}) can be handled using the fact that:
\eq
\label{doppiaderivata}
q^2\frac{d^2}{dq^2}=q\frac{d}{dq}q\frac{d}{dq}-q\frac{d}{dq}
\en
and the following identity among Eisenstein functions:
\eq
\label{e2ande4}
q\frac{d}{dq} E_2(q)=\frac{E^2_2(q)-E_4(q)}{12}
\en
which yield:
\eq
q^2\frac{d^2}{dq^2} \ln G(q)=\frac{E_4(q)-E_2^2(q)}{288}-\frac{1-E_2(q)}{24}
\label{use3}
\en
Plugging the various terms into eq.~(\ref{wholengnlo1}), one ends up with:
\eq
\label{wholengnlo2}
Z^{NLO}(R,L)=\frac{e^{-\sigma RL}}{\eta(i u)}\left\{1+\frac{\pi^2 L}{1152\sigma R^3} \left[2E_4\left(iu\right)-E_2^2\left(iu\right) \right]\right\}
\en
which is exactly eq.~(\ref{znlo}), as it was expected.

\end{document}